\documentclass{article}

\newcommand{\be}{\begin{equation}}
\newcommand{\ee}{  \end{equation}}
\newcommand{\ba}{\begin{eqnarray}}
\newcommand{\ea}{  \end{eqnarray}}
\newcommand{\ve}{\varepsilon}

\begin{document}

\title{Fourier's Law for Quasi One--Dimensional Chaotic Quantum Systems}

\author{Thomas H. Seligman, \\ Instituto de Ciencias Fisicas, \\
Universidad Nacional Autonoma de Mexico, \\ C.P. 62210 Cuernavaca,
Morelos, Mexico and \\ Centro Internacional de Ciencias, \\
C. P. 62210 Cuernavaca, Morelos, Mexico and
\\ Hans A. Weidenm{\"u}ller, \\ Max-Planck-Institut f{\"u}r
Kernphysik, \\ P.O.Box 103980, \\ 69029 Heidelberg, Germany}

\maketitle

\begin{abstract}

We derive Fourier's law for a completely coherent quasi
one--dimensional chaotic quantum system coupled locally to two heat
baths at different temperatures. We solve the master equation to first
order in the temperature difference. We show that the heat conductance
can be expressed as a thermodynamic equilibrium coefficient taken at
some intermediate temperature. We use that expression to show that for
temperatures large compared to the mean level spacing of the system,
the heat conductance is inversely proportional to the level density
and, thus, inversely proportional to the length of the system.

\end{abstract}

{\bf PACS numbers:}44.10.+l, 66.25.+g 

\section{Introduction}

In classical physics, Fourier's law states that the heat conductance
is inversely proportional to the length $L$ of a physical system. For
a system $S$ coupled at either end to two reservoirs at different
temperatures $T_1$ and $T_2$ with $T_2 > T_1$, the heat conductance
$C$ is defined by writing the heat current through the system as $C
(T_2 - T_1)$. The law $C \propto L^{-1}$ is intuitively obvious when
one thinks of $C^{-1}$ as the resistance of a macroscopic system
against heat (i.e., phonon) transport. The system can be thought of as
consisting of building blocks each with its own resistivity against
heat transport. The resistance is the sum of these resistivities and
grows linearly with $L$, resulting in $C \propto 1 / L$.

Does that law also apply when $S$ is a quantum system? The argument
just given suggests that the answer depends on the degree of coherence
of the system. Consider, for instance, a system consisting of $K$
building blocks of length $l$ each so that $L = K l$, and assume that
at the interface between neighboring blocks the system is coupled to
the outside world so that quantum coherence between neighboring blocks
is destroyed. Then the classical argument given above applies and $C$
is inversely proportional to $L$. 

Here we are interested in the heat conductance of a fully coherent
quasi one--dimensional quantum system $S$. No sources of decoherence
are present other than the decoherence due to the coupling of $S$ to
the reservoirs at either of its ends. Phonons travel coherently from
one end of $S$ to the other. Because of that coherence, the inverse
heat conductance cannot be calculated by adding the resistivities of
individual blocks in the manner described above, and it is not clear
why $C$ should be proportional to $L^{-1}$. The example of electron
transport through mesoscopic samples coupled to two external leads
actually suggests independence of $C$ on $L$. That is seen as
follows. At sufficiently low temperature, electron transport is fully
coherent~\cite{Imr02}. The electrical conductance cannot be
calculated by adding the resistivities of parts of the system. Rather,
for non--interacting electrons the conductance as given by the
Landauer--B{\"u}ttiker formula is the sum of squares of elements of
the scattering matrix, i.e., of the quantum--mechanical transition
amplitudes connecting the two leads. That formula embodies full
quantum coherence of the mesoscopic system. The elements of the
scattering matrix do not display a systematic dependence on the length
of the system. Thus, the analogy with electron transport suggests that
in a fully coherent system, $C$ should be independent of $L$. On the
other hand, numerical studies of several small quantum systems
(typically spin chains) coupled to two reservoirs strongly suggest
that $C$ is indeed inversely proportional to the length of the
system~\cite{Mic03,Mej05,Gem10}.

In the present paper we offer an analytical resolution of the
resulting paradox. We focus attention on chaotic quantum systems. We
thereby exclude both integrable systems and disordered systems with
localization. We do so because it is known that some integrable
systems do not comply with Fourier's law. Moreover, it is clear that
in disordered systems with localization, the heat conductance
decreases exponentially with length. We later identify those of our
assumptions that fail for localized systems. Starting from the master
equation which describes the coupling of a quantum system to two
reservoirs (heat baths) at different temperatures, we use a
perturbation expansion in the temperature difference to derive in
first order an expression for the heat current. We show that the heat
conductance $C$ in that expression can be written as a thermodynamic
equilibrium coefficient calculated at some intermediate temperature
$T_0$ even when the conducting system itself is not in thermodynamic
equilibrium. We show for a wide class of quantum systems that except
for a normalization factor, the resulting expression for $C$ is indeed
independent of $L$ as suggested by the analogy with mesoscopic
electron transport. The normalization factor has the form $1 / \sum_m
\exp [ - E_m / (k_B T_0) ]$ typical for equlibrium systems. Here $E_m$
are the eigenenergies of the system $S$ and $k_B$ is the Boltzmann
constant. It is easily seen that the sum over $m$ is linear in the
total level density of $S$ and, thus, grows linearly with $L$. For $C$
that yields an inverse dependence on $L$ in agreement with the results
reported in Refs.~\cite{Mic03,Mej05}.

Work on the heat conductance in quantum mechanics has a long history,
starting with Refs.~\cite{Gre52,Kub57,Lut64}, and many papers
thereafter. The length dependence of the heat conductance $C$ and the
influence of quantum coherence on the value of $C$ were discussed in
none of these early works, however. In the related
electrical--conductance problem, the importance of quantum coherence
was recognized only in the 1980s~\cite{Imr02}. During the last decade
Fourier's law in quantum mechanics has been intensely studied,
especially for quantum spin chains. In addition to the papers cited
above we mention Refs.~\cite{Bon04,Ber05,Mic06,Mic08} and the
review~\cite{Gem10} where further references may be found. The present
paper is based on a random--matrix model. The results derived within
that model and obtained as ensemble averages are generically valid for
quasi one--dimensional chaotic quantum systems.

\section{Master Equation}

We consider two heat baths (reservoirs) labeled $1$ and $2$ with
temperatures $T_1$ and $T_2 \geq T_1$ coupled to a quantum system
$S$. The Hamiltonian $H_S$ of the system has eigenfunctions $|m
\rangle$ and eigenvalues $E_m$. The occupation probabilities of the
states $| m \rangle$ are denoted by $P_m$. The stationary state of the
system $S$ is described by the master equation
\be
\sum_m (W^{(1)}_{n m}(T_1) + W^{(2)}_{n m}(T_2)) P_m = \bigg( \sum_m
(W^{(1)}_{m n}(T_1) + W^{(2)}_{m n}(T_2)) \bigg) P_n \ .
\label{1}
\ee
The derivation and conditions of validity of Eq.~(\ref{1}) have been
thoroughly discussed in the literature, see for instance
Ref.~\cite{Gar85}. To fix our notation, we give a brief derivation of
Eq.~(\ref{1}) in the Appendix. We will show that the $P_m$ are
uniquely determined and, for $T_1 \neq T_2$, in general differ from an
equilibrium distribution. We will use Eq.~(\ref{1}) to calculate the
stationary heat current and, from that, Fourier's law and an
expression for the heat conductance. In doing so we exclude systems
and/or couplings with pathological properties. Examples would be a
system consisting of two uncoupled parts, or a localized system, or a
system with coupling matrices $W^{(i)}_{m n}$ where at least one is
close to diagonal in the eigenvalue representation of $H_S$.

It is clear from the outset that within the framework of Eq.~(\ref{1})
a temperature gradient within the system $S$ cannot exist since the
temperature of the system $S$ is defined in terms of the occupation
probabilities $P_m$ and these are the same throughout the system. That
fact reflects quantum coherence throughout the system $S$. This
statement does not preclude the possibility that a suitably defined
local energy expectation value possesses a non--zero gradient,
however. For such a situation to arise, the $P_m$ must differ from the
thermodynamic equilibrium values in Eq.~(\ref{4}) below.

Explicit values of the coupling matrix elements $W^{(i)}_{m n}(T_i)$
with $i = 1,2$ are worked out in the Appendix. According to
Eq.~(\ref{37a}) these obey
\be
W^{(i)}_{n m}(T_i) = X^{(i)}_{n m} \exp \{ (\beta_i / 2) (E_m - E_n) \}
\label{2}
\ee
where $\beta_i = 1/T_i$. We put the Boltzmann constant equal to
unity. Here $X^{(i)}_{n m}$ is real, independent of temperature and,
in contrast to $W^{(i)}_{m n}$, symmetric, $X^{(i)}_{n m} = X^{(i)}_{m
n}$. Eq.~(\ref{2}) implies
\be
W^{(i)}_{n m}(T_i) = \exp [ \beta_i ( E_m - E_n ) ] W^{(i)}_{m n}(T_i)
\ .
\label{3}
\ee
Eq.~(\ref{3}) immediately yields the form of the normalized solutions
$P_n$ of Eq.~(\ref{1}) for the equilibrium case, $T_1 = T_2 = T$. With
$\beta = 1 / T$ these are given by
\be
P^{\rm eq}_n = \frac{\exp [ - \beta E_n ]}{\sum_m \exp [ - \beta
E_m ]} \ .
\label{4}
\ee

The equilibrium distribution~(\ref{4}) is independent of the values of
the elements of the coupling matrices $X^{(i)}$. Can such independence
also be expected for the non--equilibrium case $T_1 \neq T_2$? For
simplicity we take $X^{(1)} = X^{(2)} = X$ and note that the matrix
$X$ is symmetric.  Eq.~(\ref{1}) takes the form
\ba
&& \sum_m X_{n m} \bigg\{ \bigg( \exp [ (\beta_1 / 2) (E_m - E_n) ] +
\exp [ (\beta_2 / 2) (E_m - E_n) ] \bigg) P_m \nonumber \\
&& - \bigg( \exp [ (\beta_1 / 2) (E_n - E_m) ] + \exp [ (\beta_2 / 2)
(E_n - E_m) ] \bigg) P_n \bigg\} = 0 \ .
\label{4a}
\ea
For this equation to hold independently of the values of the $X_{m
n}$, each of the coefficients multiplying $X_{m n}$ must, in general,
vanish individually. It is seen immediately that that is possible only
for $T_1 = T_2$. The developments in Sections~\ref{sym} and \ref{sim}
below show that the argument is somewhat simplistic and must be
refined. Nevertheless we conclude that in the non--equilibrium case
the values of the occupation probabilities $P_m$ depend, in general,
on the values of the $X^{(i)}_{m n}$.

To determine the stationary heat current $I$ through the system $S$ we
multiply Eq.~(\ref{1}) with $E_n$ and sum over $n$. We obtain
\ba
&& \sum_{m n} E_n W^{(1)}_{n m}(T_1) P_m - \sum_{m n} E_n W^{(1)}_{m
n}(T_1) P_n \nonumber \\ && \qquad = - \sum_{m n} E_n W^{(2)}_{n
m}(T_2) P_m + \sum_{m n} E_n W^{(2)}_{m n}(T_2) P_n \ .
\label{14}
\ea
Eq.~(\ref{14}) expresses energy conservation.  We interpret the
negative left-- and the positive right--hand side of Eq.~(\ref{14}) as
the heat current $I$ (energy per unit time transferred to the system
from bath $2$ or from the system to bath $1$, respectively). That
interpretation is in keeping with the fact that the master
equation~(\ref{1}) is the stationary form of a more general equation
for the time derivative of the occupation probabilities $P_m$ of the
states $m$.

\section{Perturbation Expansion}
\label{equ}

According to Fourier's law, the heat current $I$ is proportional to
the temperature difference between the two heat baths. This suggests
using a perturbative approach in powers of $\delta T = (1/2)(T_2 -
T_1) > 0$. We accordingly expand the quantities appearing in
Eq.~(\ref{1}) around some intermediate temperature $T_0$ that obeys
$T_1 < T_0 < T_2$ in powers of $\delta T$ up to and including terms
linear in $\delta T$. The choice of $T_0$ plays an important role in
the calculation.

Expanding the coefficients $W^{(i)}(T_i)$ at $T_0$ in powers of
$\delta T = (1/2) (T_2 - T_1)$, we use
\be
\frac{\partial W^{(i)}_{n m}}{\partial T} = \frac{1}{2 T^2} (E_n -
E_m) W^{(i)}_{n m} \ ,
\label{6}
\ee
see Eq.~(\ref{2}). We likewise expand the solutions $P_m$ of
Eq.~(\ref{1}) around the equilibrium solution~(\ref{4}) at $T = T_0$,
omitting the normalization factor,
\be
P_m = \exp [ -\beta_0 E_m ] (1 + \delta P_m )
\label{7}
\ee
where $\beta_0 = 1 / T_0$. By definition of the equilibrium solution,
the terms of zeroth order in $\delta T$ in Eq.~(\ref{1}) mutually
cancel. For $i = 1,2$ we define the real symmetric matrices
\be
B^{(i)}_{m n} = \exp [ - (\beta_0 / 2) (E_m + E_n) ] X^{(i)}_{m n} \ .
\label{8}
\ee
and the vectors
\be
A^{(i)}_m = \frac{1}{T^2_0} \sum_n (E_m - E_n) B^{(i)}_{m n} \ .
\label{9}
\ee
We note that
\be
\sum_m A^{(i)}_m = 0 \ .
\label{10}
\ee
The master equation takes the form
\be
\sum_i (-)^{i} A^{(i)}_n \delta T + \sum_i \sum_m B^{(i)}_{n m}
\delta P_m - \bigg( \sum_i \sum_m B^{(i)}_{n m} \bigg) \delta P_n =
0 \ . 
\label{11}
\ee
This is a set of inhomogeneous linear equations for the unknown
quantities $\delta P_n$. The homogeneous equations possess the
non--trivial (equilibrium) solution $\delta P_n = 1$ (all $n$).
According to Eq.~(\ref{10}) the inhomogeneity is orthogonal upon that
solution. Therefore, the inhomogeneous equations possess a unique
solution $\delta P_m$ with 
\be
\sum_m \delta P_m = 0 \ .
\label{11a}
\ee
That is the solution we study in the sequel.

We consider consecutively three cases: (i) The coupling matrix
elements are equal, $X^{(1)}_{m n} = X^{(2)}_{m n}$; (ii) the coupling
matrix elements are similar, $X^{(1)}_{m n} = a X^{(2)}_{m n}$ with $a
> 0$; (iii) the coupling matrix elements are dissimilar so that neither
case (i) nor case (ii) applies.

\subsection{Symmetric Coupling}
\label{sym}

For the symmetric case with equal couplings, $X^{(1)}_{m n} =
X^{(2)}_{m n} = X_{m n}$, we choose the intermediate temperature
\be
T_0 = \frac{1}{2} \ (T_1 + T_2) \ .
\label{5}
\ee
We have $B^{(1)}_{m n} = B^{(2)}_{m n} = B_{m n}$ so that $\sum_i
(-)^{i} A^{(i)}_m = 0$ for all $m$. Therefore, the solution of
Eq.~(\ref{11}) vanishes identically, and the solution of the master
equation~(\ref{1}) is given by the equilibrium solution~(\ref{4}) with
$1 / \beta$ taken at $T = T_0$ even in the non--equilibrium case $T_1
\neq T_2$, up to and including linear terms in $\delta T$. The system
$S$ is in thermal equilibrium at the mean tempertaure $T_0$. There
cannot exist a gradient in the local expectation value of the energy.
A non--vanishing non--equilibrium solution of the master
equation~(\ref{1}) must be of second order in $\delta T$.

An explicit expression for the heat current is obtained by using for
$P_m$ the equilibrium solution (including the normalization factor)
and by expanding the left--hand side of Eq.~(\ref{14}) up to terms
linear in $\delta T$. That gives
\ba
I &=& \frac{1}{\sum_n \exp [ - \beta_0 E_n ] } \bigg( \frac{1}{2
T^2_0} \sum_{m n} (E_m - E_n)^2 B_{m n} \bigg) \ \delta T \ .
\label{15}
\ea
This, in a very general form, is Fourier's law. The heat conductance
is given by an equilibrium property of the system: The coefficient
$\sum_{m n} (E_m - E_n)^2 B_{m n}$ is calculated at the equilibrium
temperature $T_0$. The choice~(\ref{5}) of $T_0$ is justified by
hindsight: It is easy to check that for a different choice, we do not
find $\delta P_m = 0$ for all $m$. That shows the importance of the
choice of $T_0$: Apparent deviations from thermal equilibrium
indicated by nonvanishing $\delta P_m$s may only be caused by an
improper choice of $T_0$.

\subsection{Similar Couplings}
\label{sim}

We turn to case (ii) where $X^{(1)}_{m n} = a X^{(2)}_{m n}$ with $a >
0$. Again, the choice of $T_0$ is important to minimize apparent (but
unreal) deviations of the system from thermal equilibrium. Thus we
write
\be
T_0 = \alpha T_1 + (1 - \alpha) T_2
\label{21}
\ee
with $0 < \alpha < 1$ and determine $\alpha$ from the condition that
the solutions $\delta P_m$ of the linearized master
equation~(\ref{11}) vanish. That equation now takes the form
\be
[- 2 (1 - \alpha) A^{(1)}_n + 2 \alpha  A^{(2)}_n ] \delta T + \sum_i
\sum_m B^{(i)}_{n m} \delta P_m - \bigg( \sum_i \sum_m B^{(i)}_{n m}
\bigg) \delta P_n = 0 \ . 
\label{22}
\ee
Because of the definition~(\ref{9}) and the condition $X^{(1)}_{m n} =
a X^{(2)}_{m n}$ the inhomogeneity vanishes if we choose $\alpha = a /
(1 + a)$ and, thus,
\be
T_0 = \frac{1}{1 + a} \ ( a T_1 + T_2 ) \ .
\label{23}
\ee
Thus, the system $S$ is in thermal equilibrium at the temperature
$T_0$ given by Eq.~(\ref{23}). The equilibrium temperature is shifted
away from the arithmetic mean of $T_1$ and $T_2$ toward the
temperature of the heat bath with the stronger coupling to the
system. That is physically very plausible. The heat current is given
by
\ba
I &=& \frac{1}{\sum_n \exp [ - \beta_0 E_n ] } \bigg( \frac{1}
{2 T^2_0} \frac{2}{1 + a} \sum_{m n} (E_m - E_n)^2
B^{(1)}_{m n} \bigg) \ \delta T \nonumber \\
&=& \frac{1}{\sum_n \exp [ - \beta_0 E_n ] } \bigg( \frac{1}
{2 T^2_0} \frac{2a}{1 + a} \sum_{m n} (E_m - E_n)^2 B^{(2)}_{m n}
\bigg) \ \delta T \ .
\label{24}
\ea
The same conclusions as in Section~\ref{sym} apply: The system is in
thermal equilibrium at temperatute $T_0$. The heat conductance is
evaluated at the equilibrium temperature. A gradient of the local
energy expectation value and deviations from equilibrium must be of
second order in $\delta T$.

\subsection{Dissimilar Couplings}

In the case of dissimilar couplings we expect that the difference will
be rather small and statistical in nature. Indeed, the derivation of
the master equation in the Appendix is based upon the assumption that
the operators $Q^{(i)}$ that couple the system $S$ to either heat bath
are local operators that act on either end of $S$. In describing such
coupling operators explicitly, it is commonly assumed that they are
made up of local position and momentum operators. For chaotic systems
we expect the matrices $X^{(1)}$ and $X^{(2)}$ to be similar. It is
not difficult, however, to imagine systems where the matrices
$X^{(1)}$ and $X^{(2)}$ differ strongly. That is the case, for
instance, for localized systems. Then $X^{(1)}$ couples most strongly
to states localized near one end of the sample. Such states do not
couple significantly to $X^{(2)}$ at the other end, and conversely for
$X^{(2)}$. A drastic and systematic difference between the two
coupling matrices invalidates our treatment.

The case of dissimilar couplings is more complicated than the cases of
equal and similar couplings and requires some algebra. It turns out
that the occupation probabilities $P_m$ differ from an equilibrium
distribution. However, it is possibe to define a temperature $T_0$
such that the heat conductance is given by an equilibrium expression
calculated at $T_0$.

We use the ansatz Eq.~(\ref{21}) and obtain Eq.~(\ref{22}). But now
the inhomogeneity in Eq.~(\ref{22}) does not vanish for any value of
$\alpha$, and it is necessary to determine the solutions $\delta P_m
\propto \delta T$ of that equation. We define $B_{n m} = \sum_i
B^{(i)}_{n m}$ and the real symmetric matrix $\tilde{B}_{n m} = B_{n
m} - \delta_{n m} \sum_k B_{n k}$. Eq.~(\ref{22}) takes the form
\be
[- 2 (1 - \alpha) A^{(1)}_n + 2 \alpha  A^{(2)}_n ] \delta T + \sum_m
\tilde{B}_{n m} \delta P_m = 0 \ .
\label{25}
\ee
We have $\sum_m \tilde{B}_{n m} = 0$. Therefore, the matrix
$\tilde{B}_{n m}$ possesses one vanishing eigenvalue, $\lambda_1$ say,
with associated eigenvector $\{1,1,\ldots,1\}^T$. The matrix
$\tilde{B}$ can be diagonalized by a real orthogonal matrix ${\cal
O}$,
\be 
[ {\cal O} \tilde{B} {\cal O}^T ]_{m n} = \delta_{m n} \lambda_m \ .
\label{12}
\ee
The eigenvector $\{1,1,\ldots,1 \}^T$ occupies the first column of the
matrix ${\cal O}^T$, and the vector $\{ 1,1,\ldots,1 \}$ occupies the
first row of the matrix ${\cal O}$. We assume that all eigenvalues
$\lambda_m$ with $m \geq 2$ differ from zero. We multiply
Eq.~(\ref{25}) from the left with the matrix ${\cal O}$, define the
vectors $\tilde{A}^{(i)}_m = \sum_k {\cal O}_{m k} A^{(i)}_k$ and
$\delta \tilde{P}_m = \sum_k {\cal O}_{m k} \delta P_k$, and observe
that $\tilde{A}^{(i)}_1 = 0$ (see Eq.~(\ref{10})) and that
$\tilde{\delta P}_1 = 0$ (see Eq.~(\ref{11a})). For $m \geq 2$ that
yields
\be
[- 2 (1 - \alpha) \tilde{A}^{(1)}_m + 2 \alpha  \tilde{A}^{(2)}_m ]
\delta T = \lambda_m \delta \tilde{P}_m \ .
\label{13}
\ee
Solving for $\delta \tilde{P}_m$ and transforming back to $\delta P_m$
we find
\be
\delta P_m = \delta T \ \sum_n \bigg( \sum_{k \geq 2} {\cal O}_{k m}
\frac{1}{\lambda_k} {\cal O}_{k n} \bigg) [- 2 (1 - \alpha) A^{(1)}_n
+ 2 \alpha A^{(2)}_n ] \ .
\label{13a}
\ee
In order to use that result in Eq.~(\ref{7}), we normalize the
solutions defined by Eq.~(\ref{7}) so that $\sum_m P_m = 1$ and expand
the resulting expression in powers of $\delta P_m$, keeping only terms
up to first order. For $P_m$ that yields
\be
P_m = \frac{\exp [ - \beta_0 E_m ]}{\sum_n \exp [ - \beta_0 E_n ]}
\bigg( 1 + \delta P_m - \frac{\sum_k \exp [ - \beta_0  E_k ] \delta
P_k}{\sum_l \exp [ - \beta_0 E_l ]} \bigg) \ .
\label{13b}
\ee
Insertion of Eqs.~(\ref{13a}) into (\ref{13b}) gives the occupation
probabilities of the states $m$. The solutions $\delta P_m$ do not
vanish identically for any choice of $\alpha$, and there is no choice
of temperature $T_0$ for which the system would be in thermal
equilibrium. Thus, the local expectation value of the energy may not
be the same throughout the system, and it may possess a non--zero
gradient. We have not investiated that possibility.

We turn to the heat current $I$ defined in Eq.~(\ref{14}). In terms of
the solutions~(\ref{13a}) $I$ is given by
\ba
I &=& \frac{1}{\sum_n \exp [ - \beta_0 E_n ] } \bigg( \frac{\delta
T} {2 T^2_0} 2 (1 - \alpha) \sum_{m n} (E_m - E_n)^2 B^{(1)}_{m n}
\nonumber \\
&& - \sum_{m n} (E_m - E_n) B^{(1)}_{m n} \delta P_n \bigg)
\nonumber \\
&=& \frac{1}{\sum_n \exp [ - \beta_0 E_n ] } \bigg( \frac{\delta
T} {2 T^2_0} 2 \alpha \sum_{m n} (E_m - E_n)^2 B^{(2)}_{m n}
\nonumber \\
&& - \sum_{m n} (E_m - E_n) B^{(2)}_{m n} \delta P_n \bigg) \ .
\label{26}
\ea
To simplify these expressions we consider the last term in the first
of Eqs.~(\ref{26}). With the help of Eqs.~(\ref{10}) and (\ref{13a})
that term can be written as
\be
\sum_{m n} (E_m - E_n) B^{(1)}_{m n} \delta P_n = 2 T^2_0 \delta T 
\bigg( \alpha \sum_{k \geq 2} \tilde{A}^{(1)}_k \frac{1}{\lambda_k}
\tilde{A}^{(2)}_k - (1 - \alpha) \sum_{k \geq 2} \tilde{A}^{(1)}_k 
\frac{1}{\lambda_k} \tilde{A}^{(1)}_k \bigg) \ .
\label{27}
\ee
For equal couplings we have $\tilde{A}^{(1)}_k = \tilde{A}^{(2)}_k$
and the right--hand side of Eq.~(\ref{27}) vanishes for $\alpha =
1/2$. For similar couplings we have $\tilde{A}^{(1)}_k = a
\tilde{A}^{(2)}_k$ and the right--hand side of Eq.~(\ref{27})
vanishes for $\alpha = 1/(1 + a)$. In the present case of dissimilar
couplings, the expression in big round brackets is linear in $\alpha$
and, therefore, vanishes at some uniquely defined value $\alpha_1$.
That value defines via Eq.~(\ref{21}) a temperature $T^{(1)}_0$. We
expect that $T^{(1)}_0$ obeys $T_1 \leq T^{(1)}_0 \leq T_2$ and that,
therefore, $\alpha_1$ obeys $0 \leq \alpha_1 \leq 1$. Values of
$\alpha_1$ outside the interval $[0, 1]$ would be physically
implausible. Similar considerations apply to the second of
Eqs.~(\ref{26}). Again there exists a value of $\alpha_2$ and an
associated temperature $T^{(2)}_0$ for which the term linear in
$\delta P_n$ vanishes. We cannot prove, however, that $\alpha_2 =
\alpha_1$, or that $T^{(1)}_0 = T^{(2)}_0$.

Choosing $T_0 = T^{(1)}_0$ in the first and $T_0 = T^{(2)}_0$ in the
second of Eqs.~(\ref{26}) we obtain for the heat current
\ba
I &=& \frac{1}{\sum_n \exp [ - \beta^{(1)}_0 E_n ] } \bigg(
\frac{1}{(T^{(1)}_0)^2} (1 - \alpha_1) \sum_{m n} (E_m - E_n)^2
B^{(1)}_{m n}(T^{(1)}_0) \bigg) \delta T \nonumber \\
&=& \frac{1}{\sum_n \exp [ - \beta^{(2)}_0 E_n ] } \bigg(
\frac{1}{(T^{(2)}_0)^2} \alpha_2 \sum_{m n} (E_m - E_n)^2
B^{(2)}_{m n}(T^{(2)}_0) \bigg) \delta T \ .
\label{28}
\ea
In these equations, the coefficients $B^{(i)}$ are taken at the
temperatures $T^{(i)}_0$ with $i = 1,2$, and $\beta^{(i)} = 1 /
T^{(i)}_0$. We cannot show that $T^{(1)}_0 = T^{(2)}_0$ although that
would be physically most plausible. The heat conductance in both
expressions~(\ref{28}) is evaluated at some equilibrium temperature
$T^{(i)}_0$ even though the system is not in equilibrium.

\section{Length Dependence of the Heat Conductance}

In Section~\ref{equ} we have shown that in Fourier's law,
\be
I = C \ \delta T \ ,
\label{L1}
\ee
the heat conductance $C$ can always be written as an equilibrium
coefficient of the form
\be
C = \frac{\gamma}{T^2_0} \frac{1}{\sum_m  \exp [ - \beta_0 E_m ]}
\sum_{m n} (E_m - E_n)^2 \exp[ - (\beta_0 / 2)(E_m + E_n) ] X_{m n}
\ .
\label{L2}
\ee
That expression may be viewed as a special case of the Green--Kubo
formula~\cite{Gre52,Kub57}. Here $\gamma$ is a numerical coefficient
of order unity, and $T_0$ is a suitably defined temperature. The
symmetric matrix $X_{m n}$ describes the coupling of the quantum
system $S$ to one of the two heat baths. We now investigate how the
heat conductance $C$ depends on the length $L$ of the system $S$. We
consider a linear chain or a piece of wire or some other quasilinear
system, all of length $L$. We model the system in terms of random
matrices, thereby assuming that it is chaotic. We show that for any
such device, coupled at either end to heat baths with different
temperatures, the heat conductance is inversely proportional to $L$ as
is the case in classical physics, and as is found to be the case for
small spin chains~\cite{Mic03,Mej05,Gem10}.

To this end we rewrite the terms appearing in Eq.~(\ref{L2}), using
the definition~(\ref{8}),
\ba
\sum_m \exp [ - \beta_0 E_m ] &=& \int {\rm d} E \ \exp [ - \beta_0 E ]
\sum_m \delta(E - E_m) \ , \nonumber \\
\sum_{m n} (E_m - E_n)^2 B_{m n} &=& \sum_{m n} (E_m - E_n)^2 \exp[ -
\beta_0 (E_m + E_n) ] X_{m n} \nonumber \\
&=& \int {\rm d} \ve_1 \int {\rm d} \ve_2 (\ve_1 - \ve_2)^2  \exp[ -
\beta_0 (\ve_1 + \ve_2) ] \nonumber \\
&& \times 2 \pi A_0 \exp[ (\ve_1 - \ve_2)^2 / \Delta^2 \nonumber \\
&& \times \ \sum_{m n} |Q_{m n}|^2 \delta(\ve_1 - E_m) \delta(\ve_2
- E_n) \ .
\label{L3}
\ea
We have used Eqs.~(\ref{2}) and (\ref{37a}) and written the
expressions appearing in the heat conductance $C$ in terms of
length--independent energy integrals, in terms of $|Q_{m n}|^2$, and
in terms of $\delta(E - E_m)$. The constant $A_0$ measures the
strength of the coupling of the system to the heat bath, see
Eq.~(\ref{32a}). By definition the operator $Q$ couples the surface of
the system $S$ locally to the heat bath and does not depend on
$L$. Any length dependence of the terms in Eq.~(\ref{L3}) is due to
sums involving $\delta(E - E_m)$. Such sums appear differently in the
first and in the second of Eqs.~(\ref{L3}). In the first, the level
density $\sum_m \delta(E - E_m)$ itself appears as an independent
factor while in the second, the sums involve the squared matrix
element $|Q_{m n}|^2$ of $Q$. In condensed--matter physics,
expressions of similar form are referred to as the local density of
states and in nuclear physics, as the strength function of the
operator $Q$. It is obvious and confirmed below that the density of
states appearing in the first of Eqs.~(\ref{L3}) increases linearly
with the length $L$ of the system $S$. It remains to show that the
length dependence of the last term in the second of Eqs.~(\ref{L3}) is
negligible. We do so for a large class of quasi one--dimensional
systems $S$.

We exploit the fact that $Q$ is local, i.e., it acts only on one
surface of $S$. Let $P$ denote an orthogonal projector obeying $P =
P^\dag = P^2$ that projects onto a suitably chosen set of states on
that surface such that $Q_{m n} = \langle m | Q | n \rangle = \langle m
P | Q | P n \rangle$. We investigate the length dependence of the
expression
\be
\sum_m | P m \rangle \delta(E - E_m) \langle m P | = - \frac{1}{\pi} P
\ \Im \frac{1}{E^+ - H_S} \ P
\label{L4}
\ee
where $E^+$ carries an infinitesimal positive imaginary increment and
where $H_S$ is the Hamiltonian of $S$. Expression~(\ref{L4}) differs
from the expression for the total level density
\be
\rho(E) = \sum_m \delta(E - E_m) = - \frac{1}{\pi} \Im \ {\rm Trace}
\frac{1}{E^+ - H_S} \ .
\label{L5}
\ee
We show that the projector $P$ in Eq.~(\ref{L4}) has a profound
influence on the length dependence.

To model the length dependence of $S$ we think of $S$ as consisting
of $K$ blocks of fixed length $l$ each, labelled by a running index $k
= 1, \ldots, K$. The length of $S$ is then given by $K = L k$, and the
dependence on length is converted into a dependence on $K$. For $H_S$
we use a matrix representation. Let $H^{(k)}$ be the Hamiltonian
matrix for block $k$. Only neighboring blocks are coupled by
Hamiltonian matrices $W^{(k k+1)}$, and we have
\be
H_S =  \left( \matrix{ H^{(1)} & W^{(1 2)} & 0 & 0 & \ldots \cr
                       W^{(1 2)} & H^{(2)} & W^{(2 3)} & 0 & \ldots \cr
                       0 & W^{(2 3)} & H^{(3)} & W^{(3 4)} & \ldots \cr
                       0 & 0 & W^{(3 4)} & H^{(4)} & \ldots \cr
              \ldots & \ldots & \ldots & \ldots & \ldots \cr} \right)
\label{L6}
\ee
where the upper indices range from $1$ to $k$.

Let the block with $k = 1$ carry the surface on which $Q$ acts. Then
we have $P H_S P = P H^{(1)} P$. In the first block, we introduce a
complete set of orthonormal basis states $| \mu \rangle$ where $\mu =
1, 2, \ldots$.  The basis is chosen such that only the first $n$
states are surface states so that $P | \mu \rangle = 0$ for $\mu >
n$. Then $P = \sum_{\mu = 1}^n | \mu \rangle \langle \mu |$. Instead
of the strength function in Eq.~(\ref{L4}) we consider
\be
- \frac{1}{\pi} \langle 1 | \ \Im \frac{1}{E^+ - H_S} \ | 1 \rangle \ .
\label{L7}
\ee
Without loss of generality we confine ourselves here to the diagonal
element of the propagator with respect to the state with $\mu = 1$.
That choice is arbitrary, any other value of $\mu$ with $\mu \leq n$
would give the same result. We consider only diagonal elements
because the non--diagonal elements vanish on average as a consequence
of the statistical assumptions introduced below. With $E_1 = \langle 1
| H^{(1)} | 1 \rangle$ and $V_\nu = H^{(1)}_{1 \nu}$ for $\nu \geq 2$,
the matrix $H^{(1)}$ in Eq.~(\ref{L6}) is explicitly written as
\be
H^{(1)} = \left( \matrix{ E_1   & V_\rho \cr
                          V_\nu & H^{(1)}_{\nu \rho} \cr} \right) \ .
\label{L8}
\ee
Here $\nu, \rho \geq 2$. We put $E_1 = 0$ and comment on that choice 
below.

Analytical progress is possible upon introducing statistical
assumptions on the matrix elements of $H_S$. We assume that the
matrices $H^{(k)}$ with $k = 1, 2, \ldots, k$ are uncorrelated and are
members each of the Gaussian Orthogonal Ensemble (GOE) of random
matrices of dimension $N$, with the limit $N \to \infty$ eventually
taken~\cite{Guh99}. Ensemble averages are denoted by angular brackets.
The matrix elements are Gaussian--distributed random variables with
mean values zero and for $k, l, m = 1, 2, \ldots, K$ obey
\ba
\langle H^{(k)}_{\mu \nu} H^{(l)}_{\mu' \nu'} \rangle &=& (1 -
\delta_{1 k}) \delta_{k l} \frac{\lambda^2}{N} ( \delta_{\mu \nu'}
\delta_{\nu \mu'} + \delta_{\mu \mu'} \delta_{\nu \nu'}) \ ,
\nonumber \\
\langle W^{k l}_{\mu \rho} W^{k' l'}_{\mu' \rho'} \rangle &=&
\frac{w^2}{N} (\delta_ {k k'} \delta_{l l'} + \delta_{k l'}
\delta_{l k'})  ( \delta_{\mu \mu'} \delta_{\rho \rho'} +
\delta_{\mu \rho'} \delta_{\mu' \rho}) \ ,  \nonumber \\
\langle H^{(k)}_{\rho \sigma} W^{(l m)}_{\rho' \sigma'} \rangle &=&
0 \ .
\label{L9}
\ea
Because of these assumptions the average spectrum of each of the
matrices $H^{(k)}$ with $k \geq 2$ has for $N \to \infty$ the shape of
a semicircle with radius $2 \lambda$. The coupling between neighboring
blocks mediated by the matrices $W$ is characterized by the strength
parameter $w$. It is physically obvious that $w \leq \lambda$. For the
first block we introduce a special notation. We assume that the first
of Eqs.~(\ref{L9}) applies in form also to the case where $k = l = 1$
but does so only for $\mu, \nu, \mu', \nu' \geq 2$. We assume that the
$V_\mu$ in Eq.~(\ref{L8}) are Gaussian random variables with mean
value zero, not correlated with the other matrix elements, and obey
$\langle V_\mu V_\nu \rangle = v^2 \delta_{\mu \nu}$. We do so in
order to display explicitly the role of the coupling of the surface
state to the rest of the system, without any restrictions on the value
of $v^2$. For $N \to \infty$ the spectrum of $H^{(1)}$ is also of
semicircular form but the surface state plays a distinct role.

The random--matrix model introduced in Eqs.~(\ref{L9}) describes the
generic properties of chaotic quasi one--dimensional quantum systems.
It is the most general model we can think of to describe such systems.
The same model has been widely used to describe electron transport
through disordered mesoscopic samples~\cite{Mel88,Iid90}.

We work out the ensemble averages of expressions~(\ref{L5}) and
(\ref{L7}) in the framework of the random--matrix model defined in
Eqs.~(\ref{L9}). To this end we calculate the average Green function
$G(E) = \langle (E^+ - H_S)^{-1} \rangle$ of the system. Because of
the Gaussian distribution of the elements of $H_S$, that function
obeys for $N \to \infty$ the Pastur equation~\cite{Pas72}
\be
E G(E) = 1 + \langle H_S G(E) H_S \rangle G(E) \ .
\label{L10}
\ee
In view of our statistical assumptions, all non--diagonal elements of
$G(E)$ (with respect to both block index $k$ and running index $\mu$)
vanish. We write $G^{(k)}_{\mu \mu}(E)$ for the diagonal elements in
block $k$ and define $G^{(k)} = (1 / N) \sum_\mu G^{(k)}_{\mu \mu}$.
In block $1$ the sum runs from $\mu = 2$. We consider the matrix
element $G^{(1)}_{1 1}(E)$ separately because the expression in
Eq.~(\ref{L7}) is given by $(- 1 / \pi) \Im G^{(1)}_{1 1}(E)$.
According to Eq.~(\ref{L10}) we have for $N \gg 1$ and $k = 1, 2,
\ldots, K$ and with $G^{(0)} = 0 = G^{(K+1)}$
\ba
E G^{(1)}_{1 1}(E) &=& 1 + v^2 N G^{1} G^{(1)}_{1 1} \ , \nonumber \\
E G^{(k)}(E) &=& 1 + [ \lambda^2 G^{(k)} + w^2 G^{(k-1)} + w^2 G^{(k
+ 1)} ] G^{(k)} \ . 
\label{L11}
\ea
Therefore, $G^{(1)}_{1 1}(E)$ is given by
\be
G^{(1)}_{1 1}(E) = \frac{1}{E - v^2 N G^{(1)}} \ .
\label{L12} 
\ee
We observe that the factor $N v^2$ represents the total strength of
the coupling of the surface state $| 1 \rangle$ to the system
$S$. Since the state $| 1 \rangle$ couples only to states in block
$1$, that strength is independent of the length of $S$. The complex
propagator $G^{(1)}$ is obtained by solving the second set of
Eqs.~(\ref{L11}).

We solve these equations approximately. For $K \gg 1$ and $k$
somewhere in the middle of the range $[1, K]$, the form of $H_S$ in
Eq.~(\ref{L6}) suggests that $G^{(k)}$ changes slowly with $k$. We
accordingly put $G^{(k)} = G^{(k-1)} = G^{(k+1)}$. With
\be
(\lambda')^2 = \lambda^2 + 2 w^2
\label{L13}
\ee
the resulting quadratic equation for $G^{(k)}$ yields
\be
\lambda' G^{(k)}(E) = \frac{E}{2 \lambda'} - i \sqrt{1 - \bigg(
\frac{E}{2 \lambda'} \bigg)^2 } \ .
\label{L14}
\ee
The average spectrum in block $k$, proportional to the imaginary part
of $G^{(k)}$, retains the form of the semicircle. However, the range
$4 \lambda'$ of the spectrum is increased compared to the value $4
\lambda$ that applies without coupling to the neighboring blocks ($W =
0$). The increase is independent of $K$, i.e., of the length of the
system.

For blocks near the either end of the system, i.e., $k$--values close
to $1$ or $K$, the form of $H_S$ in Eq.~(\ref{L6}) and the form of
Eqs.~(\ref{L11}) suggest that the solutions retain the
form~(\ref{L14}) but with values of $\lambda'$ that are smaller than
given in Eq.~(\ref{L13}). That statement is supported by taking $K =
2$ in which case we find $(\lambda')^2 = \lambda^2 + w^2$. We conclude
that the solutions of Eqs.~(\ref{L11}) yield average spectra of
approximately semicircular shape with ranges $\lambda'$ that lie
between $\sqrt{\lambda^2 + w^2}$ and $\sqrt{\lambda^2 + 2 w^2}$ and
that are independent of $K$. An estimate of the range $\Delta E$ of
the spectrum of $H_S$ confirms that conclusion. We use that $\Delta E
\approx \sqrt{(1 / (K N)) {\rm Trace} (H_S)^2}$. For $N \gg K \gg1$ we
obtain $\Delta E \approx \sqrt{\lambda^2 + 2 w^2}$, a result that is
independent of $K$ and consistent with the values for $\lambda'$ just
mentioned.

Using the form~(\ref{L14}) for $G^{(1)}(E)$ in expression~(\ref{L7})
and putting in $G^{(1)}(E)$ the energy argument equal to zero for
simplicity we obain
\be
- \frac{1}{\pi} \Im \frac{1}{E + i v^2 N / \lambda'} \ .  
\label{L15}
\ee
The width $2 v^2 N / \lambda$ of the strength function of the state $|
1 \rangle$ is essentially given by the total strength $N v^2$ of the
coupling of that state to the system $S$ divided by the range of the
spectrum of $S$. That is a physically very plausible result. Neither
$N v^2$ nor $\lambda'$ depend on the length of the system. The
result~(\ref{L15}) can easily be extended by taking into account the
full form of the propagator in Eq.~(\ref{L14}) and by dropping the
assumption $E_1 = 0$. Instead of the form~(\ref{L15}) one obtains an
expression of Breit--Wigner form centered at $E_1$ with a width and a
level shift due to the imaginary and the real parts of $G^{(1)}$,
respectively. That expression is also independent of $K$. The same
statement obviously applies to the full expression~(\ref{L4}).

We compare this result with the total level density of $S$ given by
Eq.~(\ref{L5}). From the definition of $G^{(k)}$ it follows that
\be
\langle \rho(E) \rangle = - \frac{N}{\pi} \Im \sum_k G^{(k)}(E)
\label{L16}
\ee
and, with $G^{(k)}(E)$ almost independent of $k$, $\langle \rho(E)
\rangle \propto K \propto L$. We conclude that the double sum in the
numerator of Eq.~(\ref{L2}) is independent of $L$ while the single sum
in the denominator is linear in $L$. As a result the heat conductance
$C$ is inversely proportional to $L$.

\section{Summary and Conclusions}

For a completely coherent chaotic quasi one--dimensional quantum
system coupled to two heat baths at different temperatures, we have
solved the master equation up to first order in the temperature
difference and obtained Fourier's law. For equal or similar couplings
at both ends of the system we have shown that the heat conductance $C$
in Fourier's law can always be written as an equilibrium
coefficient. In the case of dissimilar couplings the same statement
holds generically for chaotic systems but not, for instance, for
localized systems.

We have used that result to investigate the dependence of $C$ on the
length $L$ of the quantum system. Intuitive arguments based on quantum
coherence and the analogy with electron transport through mesoscopic
systems both suggest that $C$ be independent of $L$, in contrast to
numerical evidence~\cite{Mic03,Mej05} showing that $C \propto L^{-1}$.
We have resolved that discrepancy by showing that aside from a
normalization factor, $C$ is indeed independent of $L$. The entire
length dependence of $C$ is found to be due to the normalization
factor and determined by the density of states. The latter increases
linearly with $L$ and yields $C \propto L^{-1}$.

The length dependence of the remaining term in $C$ is determined by
that of the last factor in Eq.~(\ref{L3}). That factor bears a close
analogy to the spreading width in nuclear physics and to the local
density of states in condensed--matter physics. In both cases, it is
known that the values are not affected when the dimension of the
system is increased. In the present case, we have used a
random--matrix approach to model the length dependence. That approach
yields generic results~\cite{Guh99} (exceptions have measure zero with
respect to the probability measure that defines the random--matrix
ensemble). We have also used the essential fact that the coupling to
the heat baths is local and linked to the surface. We have shown that
that fact guarantees the length--independence of the relevant term in
Eq.~(\ref{L3}).

Having resolved the discrepancy, we may turn the question around and
ask: Why does electron transport in mesoscopic systems not show a
similar length dependence due to a normalization factor? The answer
lies in the temperatures at which heat transport and electron
transport are considered. To insure quantum coherence, electron
transport is experimentally studied close to zero Kelvin. At such low
temperature, only the lowest eigenvalues $E_m$ of the system would
contribute to $C$. The random--matrix model we use is viable only when
the density of states is sufficiently high, i.e., when the temperature
$T_0$ at which $C$ is evaluated, is very much larger than the mean
level spacing near the ground state. The random--matrix model is not
expected to account correctly for effects near the end points of the
spectrum.

Although $C$ is inversely proportional to $L$ in both classical and
quantum physics, the causes for that dependence are seen to be
strikingly different. In classical physics and for systems of
macroscopic size, lack of quantum coherence causes the total
resistance to be the sum of the resistivities of subsystems and, thus,
$C \propto L^{-1}$. For a fully coherent quantum system, the inverse
length dependence of $C$ is due to the linear increase with $L$ of the
level density of the system.

{\bf Acknowledgements.} We thank F. Leyvraz, T. Prosen, and
M. Znidaric for useful discussions, and D. Huse for correspondence.
THS acknowledges financial support under the projects IN114310 by
PAPIIT, Universidad Nacional Autonoma de Mexico and 79613 by CONACyT.

\section*{Appendix: Derivation of the Master Equation}

In Ref.~\cite{Lut99}, the master equation for a quantum system coupled
to a single heat bath was derived. Here we use the same derivation for
a quantum system coupled to two heat baths with temperatures $T_1$ and
$T_2$ to obtain Eq.~(\ref{1}). We only sketch those parts of the
derivation that differ from Ref.~\cite{Lut99} and do not repeat here
the discussion of the conditions under which the derivation holds.
These are the same as in Ref.~\cite{Lut99}.

The quantum system $S$ has Hamiltonian $H_S$, eigenvalues $E_m$, and
eigenfunctions $| m \rangle$, with $m = 1, 2, \ldots$. With $i = 1,2$
the two baths have Hamiltonians $H_i$, eigenvalues $\ve^{(i)}_a$ and
eigenfunctions $| i a \rangle$, respectively, with $a = 1, 2, \ldots$.
The coupling between each of the baths $i = 1,2$ and the system has
the form $W^{(i)} = Q^{(i)} V^{(i)}$.  Here $Q^{(i)}$ are local
operators that act on either end of the system $S$ while the
$V^{(i)}_{a b}$ are uncorrelated random matrices to be specified
below. The total Hamiltonian is
\ba
{\cal H} &=& H_S + H_1 + H_2 + \sum_i W^{(i)} \nonumber \\
&=& H_0 + W \ .
\label{31}
\ea
To define the ensemble we consider case (II) of Ref.~\cite{Lut99} and
assume that the $W^{(i)}$ are Gaussian--distributed random variables
with zero mean values and second moments given by
\ba
\overline{\langle m ia \ | \ W^{(i)} \ | \ n ib  \rangle \langle p ic
\ | \ W^{(i)} \ | \ q id \rangle} &=& [ \delta_{m p} \delta_{n q}
\delta_{a c} \delta_{b d} + \delta_{m q} \delta_{n p} \delta_{a d}
\delta_{b c} ] \nonumber \\
&& \qquad \times |Q^{(i)}_{m n}|^2 \overline{(V^{(i)}_{a b})^2} \ ,
\ i = 1,2 \ ,
\nonumber \\ 
\overline{\langle 1a m | W^{(1)} | 1b n \rangle \langle 2c p | W^{(2)}
| 2d q \rangle} &=& 0 \ . 
\label{32}
\ea
The overbar denotes the average over the ensemble. The second of
Eqs.~(\ref{32}) shows that $W^{(1)}$ and $W^{(2)}$ are uncorrelated.
In the first of Eqs.~(\ref{32}) we assume standard random--matrix
properties. We assume that the coupling of the system $S$ to the heat
bath described by the matrices $Q^{(i)}$ acts only on the (left or
right) surface of the system and not on its volume. For the matrices
$V^{(i)}$ we assume as in Ref.~\cite{Lut99} that with $i = 1, 2$
\be
\overline{[V^{(i)}_{a b}]^2} = A^{(i)}_0 [\rho^{(i)}(\ve_a)
\rho^{(i)}(\ve_b)]^{- 1/2} \exp [ - (\ve^{(i)}_a - \ve^{(i)}_b)^2 /
(2 \Delta^2_i) ]
\label{32a}
\ee
where $A^{(i)}_0$ and $\Delta_i$ are constants and where
\be
\rho^{(i)}(\ve) \approx \rho^{(i)}_0 \exp ( \beta_i \ve )
\label{32b}
\ee
is the level density and $\beta_i = 1 / T_i$ the inverse temperature
of the heat bath labelled $(i)$.

The sum $\sum_i W^{(i)}$ is also a Gaussian random variable with mean
value zero. This fact suffices to derive for the average density
matrix $\overline{\rho(t, t')}$ of the total system governed by ${\cal
H}$ the integral equation
\be
\overline{\rho(t, t')} = \overline{U(t)} \rho(0, 0) \overline{U^\dag(t)}
+ \int_0^t {\rm d} \tau \int_0^{t'} {\rm d} \tau' \ \overline{U(t -
\tau)} \ \overline{W \overline{\rho(\tau, \tau')} W} \
\overline{U^\dag(t' - \tau')} \ .   
\label{33}
\ee
The averaged time--evolution operator obeys
\ba
\overline{U(t)}  &=& \exp [ - i H_0 t ] - \int_)^t {\rm d} t_1
\int_0^{t_1} {\rm d} t_2 \ \exp [ - i H_0 (t - t_1) ] \nonumber \\
&& \times \overline{ W \exp [ - i H_0 (t_1 - t_2 ) ] W} \
\overline{U(t_2)} \ . 
\label{34}
\ea
With the help of the assumptions~(\ref{32}) it is easily seen that
\ba
&& \langle m 1a 2b \ | \ \overline{U(t)} \ | \ n 1c 2d \rangle
\nonumber \\
&& = \delta_{m n} \delta_{a c} \delta_{b d} \exp [ - i (E_m +
\ve^{(1)}_a + \ve^{(2)}_b) t - (1/2)( \Gamma^{(1)}_{m a} +
\Gamma^{(2)}_{m b}) t ] \Theta(t) 
\label{35}
\ea
where $\Theta(t)$ is the Heaviside function and where 
\be
\Gamma^{(i)}_{m a} = 2 \pi \sum_{n b} \overline{ \langle m ia |
W^{(i)} | n ib \rangle^2} \ \delta(E_m + \ve^{(i)}_a - E_n -
\ve^{(i)}_b) \ , \ i = 1,2 \ .
\label{36}
\ee
In Eq.~(\ref{35}) the sum of two $\Gamma$s appears because $W^{(1)}$
and $W^{(2)}$ are uncorrelated and may cause different damping.

We take the trace of Eq.~(\ref{33}) with respect to both heat baths
and obtain
\ba
&& \sum_{a b} \langle m 1a 2b \ | \ \overline{\rho(t, t')} \ | \ n
1a 2b \rangle = \sum_{a b} \exp [ - i (E_m t - E_n t') ]
\nonumber \\
&& \qquad \times \exp [ - i (\ve^{(1)}_a + \ve^{(2)}_b) (t - t') ]
\nonumber \\
&& \qquad \times \exp [ - (1/2) (\Gamma^{(1)}_{m a} + \Gamma^{(2)}_{m
b}) t - (1/2)(\Gamma^{(1)}_{n a} + \Gamma^{(2)}_{n b})t' ]
\nonumber \\
&& \qquad \times \langle m 1a 2b \ | \ \rho(0, 0) \ | \ n 1a 2b
\rangle
\nonumber \\
&& + \delta_{m n} \sum_{a b} \int_0^t {\rm d} \tau \int_0^{t'} {\rm d}
\tau' \exp [ - i (E_m + \ve^{(1)}_a + \ve^{(2)}_b)(t - \tau) ]
\nonumber \\
&& \qquad \times \exp [ - (1/2) ( \Gamma^{(1)}_{m a} + \Gamma^{(2)}_{m
b})(t - \tau) ]
\nonumber \\
&& \qquad \times \bigg( \sum_{n c} \overline{|\langle m 1a 2b \ | \
W^{(1)} \ | \ n 1c 2b \rangle |^2} \langle n 1c 2b \ | \
\overline{\rho(\tau, \tau')} \ | \ n 1c 2b \rangle
\nonumber \\
&& \qquad + \sum_{n d} \overline{|\langle m 1a 2b \ | \ W^{(2)} \ | \
n 1a 2d \rangle |^2} \langle n 1a 2d \ | \ \overline{\rho(\tau, \tau')}
\ | \ n 1a 2d \rangle \bigg)
\nonumber \\ 
&& \qquad \times \exp [ \ i (E_m + \ve^{(1)}_a + \ve^{(2)}_b)(t' -
\tau') ]
\nonumber \\
&& \qquad \times \exp [ - (1/2) ( \Gamma^{(1)}_{m a} + \Gamma^{(2)}_{m
b})(t' - \tau') ] \ . 
\label{37}
\ea
The Kronecker delta on the right--hand side of Eq.~(\ref{37}) follows
from the first of Eqs.~(\ref{32}). Eq.~(\ref{37}) shows that the
nondiagonal elements of the reduced density matrix of system $S$ decay
exponentially in time. Therefore, we focus attention on the diagonal
elements with $m = n$, for $t = t'$ denoted by $P_n(t)$. We
differentiate Eq.~(\ref{37}) with respect to $t$ and $t'$. In the
resulting gain terms we use the weak--coupling assumption and the
resulting Markov approximation. We make use of Eqs.~(\ref{32a}) and
(\ref{32b}). With
\be
W^{(i)}_{n m}(T_i) = 2 \pi A^{(i)}_0 |Q^{(i)}_{n m}|^2 \exp [ - (E_n
- E_m)^2 / ( 2 \Delta^2_i ) ] \exp [ (1 / 2) \beta_i (E_m - E_n) ]
\label{37a}
\ee
that yields Eqs.~(\ref{1}) and (\ref{2}).

\end{document}